\documentclass{aa}
\usepackage{graphicx}
\usepackage{natbib}
\newcommand\eg{{\it e.g.} }
\newcommand\etal{et~al.}
\def\spose#1{\hbox to 0pt{#1\hss}}
\newcommand\simlt{\mathrel{\spose{\lower 3pt\hbox{$\mathchar"218$}}
     \raise 2.0pt\hbox{$\mathchar"13C$}}}
\newcommand\lesssim{\mathrel{\spose{\lower 3pt\hbox{$\mathchar"218$}}
     \raise 2.0pt\hbox{$\mathchar"13C$}}}
\newcommand\simgt{\mathrel{\spose{\lower 3pt\hbox{$\mathchar"218$}}
     \raise 2.0pt\hbox{$\mathchar"13E$}}}
\newcommand\gtrsim{\mathrel{\spose{\lower 3pt\hbox{$\mathchar"218$}}
     \raise 2.0pt\hbox{$\mathchar"13E$}}}
\begin{document}
\title{A Sample of Ultra Steep Spectrum Sources Selected from the Westerbork In the Southern Hemisphere (WISH) survey.}
\titlerunning{USS sources from the WISH survey}

\author{Carlos De Breuck\inst{1}\thanks{Marie Curie fellow} \and Yuan Tang\inst{2} \and A.\ G.\ de Bruyn\inst{2,3} \and Huub R\"ottgering\inst{4} \and Wil van Breugel\inst{5}}

\offprints{Carlos De Breuck}
\institute{Institut d'Astrophysique de Paris, 98bis Boulevard Arago, 75014 Paris, France\\ \email{debreuck@iap.fr} \and ASTRON, Postbus 2, 7990 AA Dwingeloo, The Netherlands \\ \email{tang,ger@nfra.nl} \and Kapteyn Astronomical Institute, PO Box 800, 9700 AV Groningen, The Netherlands \and Sterrewacht Leiden, Postbus 9513, 2300 RA Leiden, The Netherlands\\ \email{rottgeri@strw.leidenuniv.nl} \and IGPP/LLNL, L-413, 7000 East Ave, Livermore, CA 94550, USA\\ \email{wil@igpp.ucllnl.org}}

\date{Received 2002 June 25; accepted 2002 August 1}

\abstract{
The 352~MHz Westerbork In the Southern Hemisphere (WISH) survey is the southern extension of the WENSS, covering 1.60~sr between $-9\degr < \delta < -26\degr$ to a limiting flux density of $\sim$18~mJy ($5\sigma$). Due to the very low elevation of the observations, the survey has a much lower resolution in declination than in right ascension ($54\arcsec \times 54\arcsec {\rm cosec}\delta$). A correlation with the 1.4~GHz NVSS shows that the positional accuracy is less constrained in declination than in right ascension, but there is no significant systematic error.
We present a source list containing 73570 sources.
We correlate this WISH catalogue with the NVSS to construct a sample of faint Ultra Steep Spectrum (USS) sources, which is accessible for follow-up studies with large optical telescopes in the southern hemisphere. This sample is aimed at increasing the number of known high redshift radio galaxies to allow detailed follow-up studies of these massive galaxies and their environments in the early Universe.
\keywords{surveys -- radio continuum: general -- radio continuum: galaxies -- galaxies: active}
}

\maketitle
%

\section{Introduction}
Powerful radio sources provide excellent targets to probe the formation and evolution of galaxies out to cosmological distances. The Hubble $K-z$ diagram of radio and near$-$IR selected galaxies shows that at $z \simgt 1$, the host galaxies of powerful radio sources are $>$2 magnitudes brighter than HDF field galaxies \citep{deb02}. Because there are strong arguments that this $K-$band emission is due to starlight, and not due to direct or scattered AGN contributions, high redshift radio galaxies (HzRGs) are among the most massive galaxies known at high redshift.
This is consistent with the observations at low redshifts ($z \simlt 1$), where radio galaxies are uniquely identified with massive ellipticals \citep[\eg][]{bes98,mcl00}.
Because HzRGs pinpoint over-dense regions in the early Universe, they have also been successfully used as tracers of proto-clusters at very high redshifts \citep[\eg][]{lef96,pas96,pen00,ven02}.
\begin{figure*}[ht]
\centering
\includegraphics[width=9cm,angle=-90]{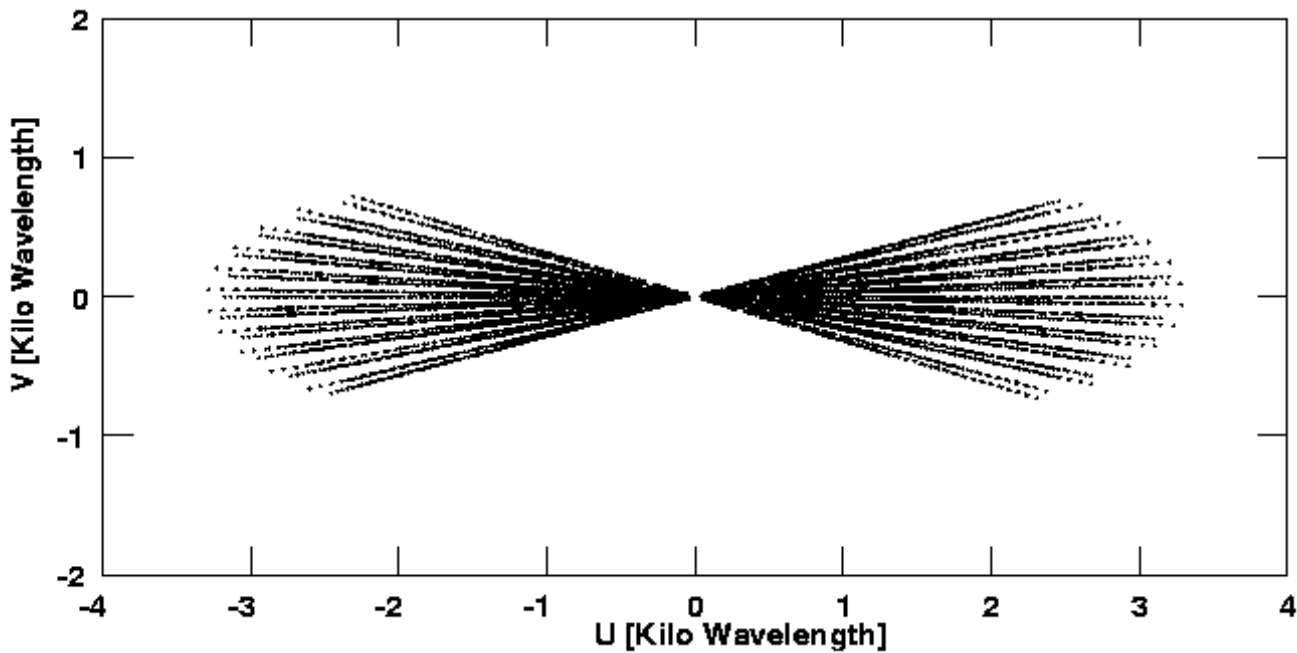}
\caption{Sample $u-v$ coverage plot of the field centered at $\alpha=1^h28^m, \delta=-20^{\circ}$. The $v-$coverage is limited due to the low declination. Note the excellent radial coverage due the the bandwidth synthesis technique.}
\label{sampleUV}
\end{figure*}

The first redshift surveys of radio sources targeted only the brightest objects in the sky \citep[3CR][]{ben62,spi85}. Present-day surveys reach flux densities several orders of magnitude fainter than the 3CR \citep[\eg][]{mcc96,lac99,raw01,wil02}.
However, because the optical spectroscopy of the host galaxies requires substantial integration times on 3$-$10m class telescopes, these flux density limited surveys necessarily need to be limited in sky area. This excludes the objects with the lowest space density, such as the most distant luminous radio galaxies \citep[\eg][]{blu98,jar01b}. To find such objects, additional 'high redshift filters' need to be applied to the samples of radio surveys, at the expense of completeness. The most efficient filter is the selection of sources with ultra steep radio spectra (USS; $\alpha \lesssim -1; S \propto \nu^{\alpha}$). The success of this USS technique is mainly based on the $k-$correction of the generally concave radio spectrum of powerful radio galaxies. Several such USS samples have shown to be much more efficient in finding $z>2$ radio galaxies than the complete samples \citep[\eg][]{rot97,ste99,deb01,jar01a}.

It is now possible to define large, well defined samples of USS sources using a new generation of large area radio surveys: the Westerbork Northern Sky Survey \citep[WENSS; 325~MHz][]{ren97}, the Texas survey \citep[365~MHz]{dou96}, The Sydney University Molonglo Sky Survey \citep[SUMSS][]{boc99}, the NRAO VLA Sky Survey \citep[NVSS; 1.4~GHz][]{con98}, and the Faint Images of the Radio Sky at Twenty centimeters \citep[FIRST; 1.4~GHz][]{bec95}.
\citet{deb00} have used these surveys to define a sample of 669 USS sources covering the entire sky outside the Galactic plane. However, their samples necessarily favour the northern hemisphere, because the WENSS survey, which is an order of magnitude deeper than the Texas survey, covers only the sky at $\delta > +29$\degr. In this paper, we introduce the Westerbork In the Southern Hemisphere (WISH) survey, the southern extension of the WENSS. We use WISH in combination with NVSS to define a fainter sample of USS sample in the $-9\degr< \delta < -26\degr$ region, in analogy with the northern WENSS$-$NVSS sample of \citet{deb00}. The construction of such a southern hemisphere sample is especially timely due to the advent of several 8m class telescopes in the southern hemisphere, which can be used for the optical/near$-$IR identification and spectroscopy of the host galaxies. 

The layout of this paper is as follows. In \S 2, we introduce the WISH survey, and compare the data products with the WENSS. In \S 3, we define the WISH$-$NVSS USS sample. \S 4 compares this new sample with previous samples and \S 5 concludes with an overview of the planned observations of this sample.

\section{The WISH survey}
\subsection{Motivation}
The 325~MHz WENSS survey \citep{ren97} has proved to be an extremely valuable tool for extra-galactic and galactic Astronomy. Some of the scientific applications include (i) the study of large-scale structure using radio sources \citep{ren99a}, (ii) the search for high redshift radio galaxies \citep{deb00}, (iii) the selection of faint Gigahertz Peaked Spectrum sources \citep{sne98}, (iv) the selection of Giant Radio Galaxies \citep{sch01}, (v) the construction of samples to search for gravitational lenses in the context of the Cosmic Lens All Sky Survey \citep{mey95}, (vi) the selection of optically bright galaxies with radio counterparts to construct the local radio luminosity functions of elliptical and spiral galaxies \citep{der98}, (vii) the study of linear polarization of the diffuse galactic radio background \citep{hav00}, and (viii) the study of pulsars \citep{kou00}.

The WENSS survey is the most sensitive low-frequency survey covering a substantial fraction (25 \%) of the sky. The limitation of the sky coverage is due to the East-West construction of the WSRT.  
With the advent of several new large optical facilities in the southern hemisphere (VLT, Gemini South, Magellan), there is a clear need for equally sensitive samples in the South.  
Because the WSRT can also observe a part of the southern sky, we extended the WENSS by carrying out a similar Westerbork survey in the southern hemisphere, the WISH.
\begin{figure*}[ht]
\centering
\includegraphics[width=17cm,angle=-90]{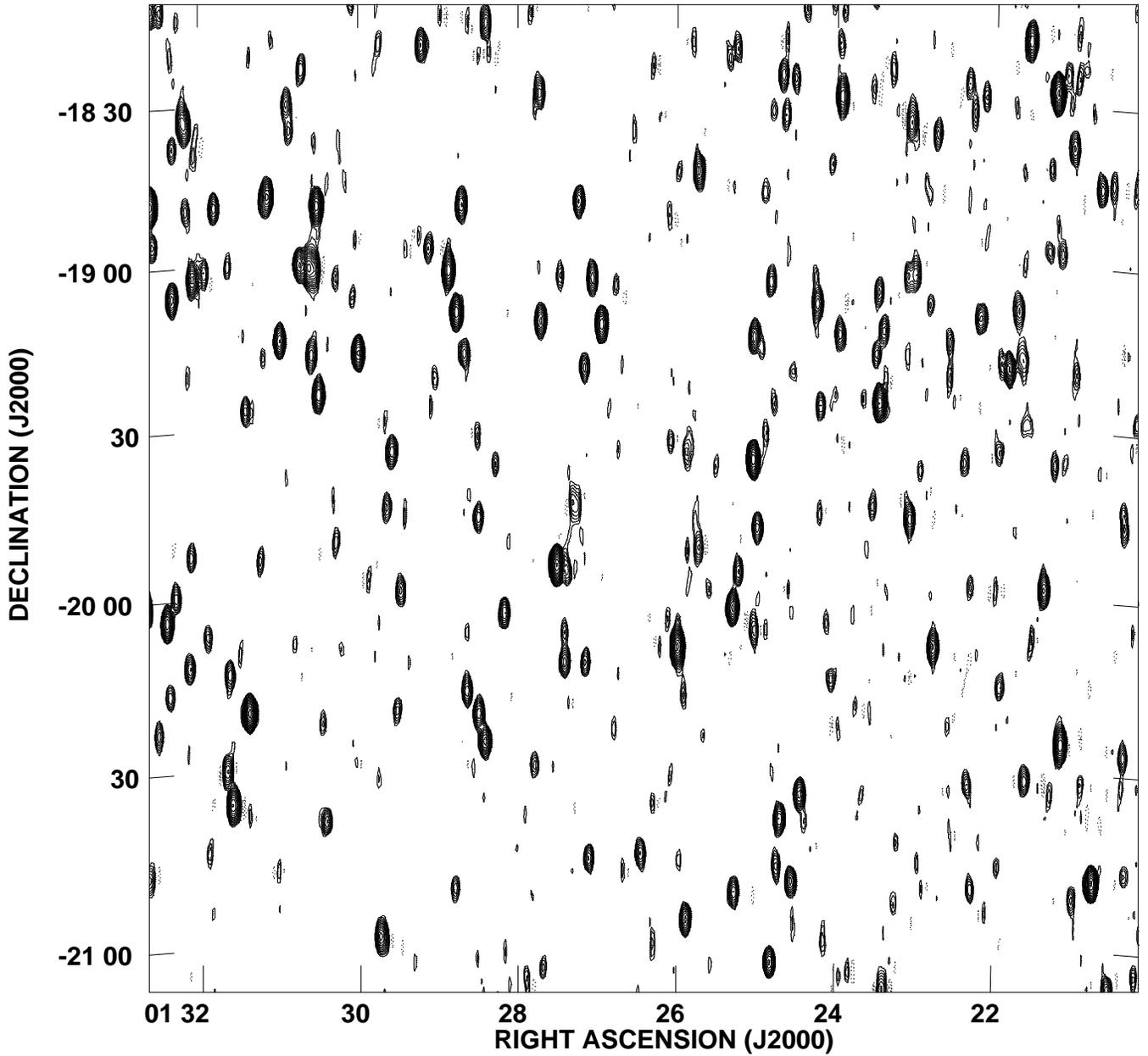}
\caption{Central part of the field centered at $\alpha=1^h28^m, \delta=-20^{\circ}$. The contour scheme is a geometric progression in $\sqrt 2$, which implies a factor 2 change in surface brightness every two contours (negative contours are dotted). The first contour is at 3$\sigma$ with $\sigma$=2.0~mJy~Beam$^{-1}$. Note that even with the elongated beam, we can still resolve several sources.}
\label{sampleimage}
\end{figure*}

\subsection{Survey design}
The main goal of the WISH is to cover as large as possible an area that can be observed with the VLT. The limitations are the latitude of Westerbork, where the horizon is at $\delta = -37$\degr, and an infinite elongation of the synthesized beam towards $\delta = 0$\degr. We therefore imposed a northern limit of $\delta < -9$\degr, to obtain a beam with a ratio of the major to minor axis of $<$6. In order to obtain at least 4 hours of hour angle coverage, we limited the survey to $\delta \simgt -26$\degr. However, the uv-coverage is still limited (see Fig.~\ref{sampleUV}), which results in a synthesized beam with large near-in sidelobes. To limit the effects of this problem, we decided to avoid the Galactic Plane ($|b|>15$\degr) with its bright extended emission. 

Apart from the above limitations, the design of WISH is based on that of WENSS. We used the broadband back-end with 8 bands of 5 MHz to improve the uv-coverage using bandwidth synthesis. The central frequencies of these 8 bands are 325.0, 333.0, 341.0, 347.0, 354.85, 366.6, 371.3, and 377.3~MHz. This bandwidth synthesis mosaicing technique was used before for the WENSS polar cap area ($\delta > +75$\degr) and proved very successful \citep{ren99b}. Because this requires only three configurations of the $9-A$ baseline, instead of six for WENSS, WISH was also more efficient in observing time.

WISH consists of mosaics of $8 \times 8$ pointings each.  With a grid-step of 1.25$^\circ$ each mosaic covers $10\degr \times 10\degr = 100$ square degrees (10 degrees in Declination and 42 minutes in Right Ascension).  Each pointing was observed for 20~sec. With a move time of 10~s we have thus covered a mosaic once every 32~minutes.  In an average observing time of 6~hours we have therefore obtained 11 cuts at each position. By observing in three different configurations (9A=48m, 72m and 96m) we have thus accumulated 33 cuts. 

\subsection{Observations and data reduction}
The observations for WISH started in the Autumn 1997, and were concluded in the Spring of 1998. The observational techniques, data reduction, and source extraction process are identical to the polar cap region of the WENSS, and are discussed by \citet{ren99b}. Table~\ref{frames} lists the 49 frames of the WISH with their respective field centers. Figure~\ref{sampleUV} shows the $u-v$ plane coverage of a sample field, and Fig.~\ref{sampleimage} the central $3\degr \times 3\degr$ of this field. Although the limited $v-$coverage leads to beam shapes elongated along the North-South direction, we can still resolve several sources into individual components.

\begin{figure*}
\centering
\includegraphics[width=17cm]{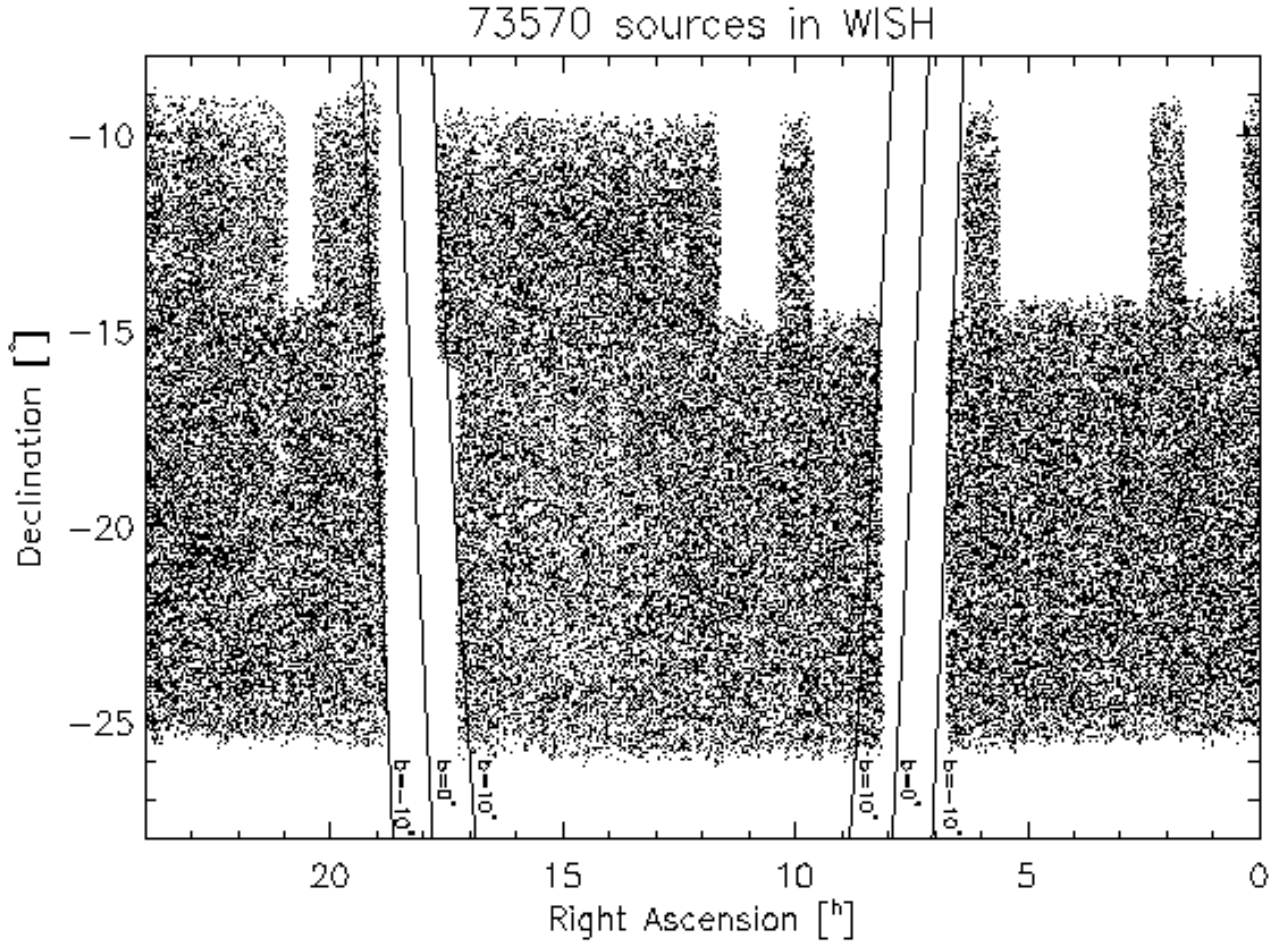}
\caption{Location of the 73570 sources detected in WISH. The galactic plane ($|b|<10$\degr) was not observed.}
\label{layout}
\end{figure*}

The total sky area of WISH is 1.60~sr (half the WENSS area). Figure~\ref{layout} shows the total sky coverage of WISH. Note that $\sim$38\% of the $\delta=-13\degr$ fields could not be observed due to the limited observing time.

\begin{table}
\caption{The 49 frames of the WISH. $C$ is the flux density correction factor discussed in \S 2.4.3.}
\label{frames}
\begin{center}
\begin{tabular}{lccc}\hline
Frame & \multicolumn{2}{c}{Mosaic center (B1950)} & $C$ \\
      & RA & Dec & \\ \hline
SNH13\_029 & $01^h56^m00^s$ & $-$13\degr00\arcmin & 0.855\\
SNH13\_089 & $05^h56^m00^s$ & $-$13\degr00\arcmin & 0.809\\
SNH13\_149 & $09^h56^m00^s$ & $-$13\degr00\arcmin & 0.830\\
SNH13\_179 & $11^h56^m00^s$ & $-$13\degr00\arcmin & 0.835\\
SNH13\_189 & $12^h36^m00^s$ & $-$13\degr00\arcmin & 0.823\\
SNH13\_199 & $13^h16^m00^s$ & $-$13\degr00\arcmin & 0.807\\
SNH13\_209 & $13^h56^m00^s$ & $-$13\degr00\arcmin & 0.838\\
SNH13\_219 & $14^h36^m00^s$ & $-$13\degr00\arcmin & 0.860\\
SNH13\_229 & $15^h16^m00^s$ & $-$13\degr00\arcmin & 0.884\\
SNH13\_239 & $15^h56^m00^s$ & $-$13\degr00\arcmin & 0.846\\
SNH13\_249 & $16^h36^m00^s$ & $-$13\degr00\arcmin & 0.879\\
SNH13\_259 & $17^h16^m00^s$ & $-$13\degr00\arcmin & 0.831\\
SNH13\_289 & $19^h16^m00^s$ & $-$13\degr00\arcmin & 0.992\\
SNH13\_299 & $19^h56^m00^s$ & $-$13\degr00\arcmin & 0.903\\
SNH13\_319 & $21^h16^m00^s$ & $-$13\degr00\arcmin & 0.833\\
SNH13\_329 & $21^h56^m00^s$ & $-$13\degr00\arcmin & 0.822\\
SNH13\_339 & $22^h36^m00^s$ & $-$13\degr00\arcmin & 0.853\\
SNH13\_349 & $23^h16^m00^s$ & $-$13\degr00\arcmin & 0.858\\
SNH13\_359 & $23^h56^m00^s$ & $-$13\degr00\arcmin & 0.850\\
SNH20\_000 & $00^h00^m00^s$ & $-$20\degr00\arcmin & 0.818\\
SNH20\_011 & $00^h44^m00^s$ & $-$20\degr00\arcmin & 0.802\\
SNH20\_021 & $01^h24^m00^s$ & $-$20\degr00\arcmin & 0.816\\
SNH20\_032 & $02^h08^m00^s$ & $-$20\degr00\arcmin & 0.790\\
SNH20\_042 & $02^h48^m00^s$ & $-$20\degr00\arcmin & 0.755\\
SNH20\_053 & $03^h32^m00^s$ & $-$20\degr00\arcmin & 0.709\\
SNH20\_063 & $04^h12^m00^s$ & $-$20\degr00\arcmin & 0.691\\
SNH20\_074 & $04^h56^m00^s$ & $-$20\degr00\arcmin & 0.880\\
SNH20\_084 & $05^h36^m00^s$ & $-$20\degr00\arcmin & 0.855\\
SNH20\_095 & $06^h20^m00^s$ & $-$20\degr00\arcmin & 0.797\\
SNH20\_127 & $08^h28^m00^s$ & $-$20\degr00\arcmin & 0.811\\
SNH20\_137 & $09^h08^m00^s$ & $-$20\degr00\arcmin & 0.781\\
SNH20\_148 & $09^h52^m00^s$ & $-$20\degr00\arcmin & 0.803\\
SNH20\_158 & $10^h32^m00^s$ & $-$20\degr00\arcmin & 0.761\\
SNH20\_169 & $11^h16^m00^s$ & $-$20\degr00\arcmin & 0.716\\
SNH20\_179 & $11^h56^m00^s$ & $-$20\degr00\arcmin & 0.858\\
SNH20\_190 & $12^h40^m00^s$ & $-$20\degr00\arcmin & 0.870\\
SNH20\_200 & $13^h20^m00^s$ & $-$20\degr00\arcmin & 0.898\\
SNH20\_211 & $14^h04^m00^s$ & $-$20\degr00\arcmin & 1.039\\
SNH20\_221 & $14^h44^m00^s$ & $-$20\degr00\arcmin & 1.063\\
SNH20\_232 & $15^h28^m00^s$ & $-$20\degr00\arcmin & 1.128\\
SNH20\_242 & $16^h08^m00^s$ & $-$20\degr00\arcmin & 1.087\\
SNH20\_253 & $16^h52^m00^s$ & $-$20\degr00\arcmin & 1.122\\
SNH20\_287 & $19^h08^m00^s$ & $-$20\degr00\arcmin & 1.199\\
SNH20\_297 & $19^h48^m00^s$ & $-$20\degr00\arcmin & 0.974\\
SNH20\_308 & $20^h32^m00^s$ & $-$20\degr00\arcmin & 0.846\\
SNH20\_318 & $21^h12^m00^s$ & $-$20\degr00\arcmin & 0.937\\
SNH20\_329 & $21^h56^m00^s$ & $-$20\degr00\arcmin & 0.785\\
SNH20\_339 & $22^h36^m00^s$ & $-$20\degr00\arcmin & 0.904\\
SNH20\_350 & $23^h20^m00^s$ & $-$20\degr00\arcmin & 0.846\\
\end{tabular}
\end{center}
\end{table}

The final WISH catalogue contains 73570 sources, and is available on the WENSS homepage (note that this catalogue has already been corrected for the flux density problem described in \S 2.4): http://www.strw.leidenuniv.nl/wenss

\subsubsection{Noise}
\begin{figure}
\centering
\includegraphics[width=8cm]{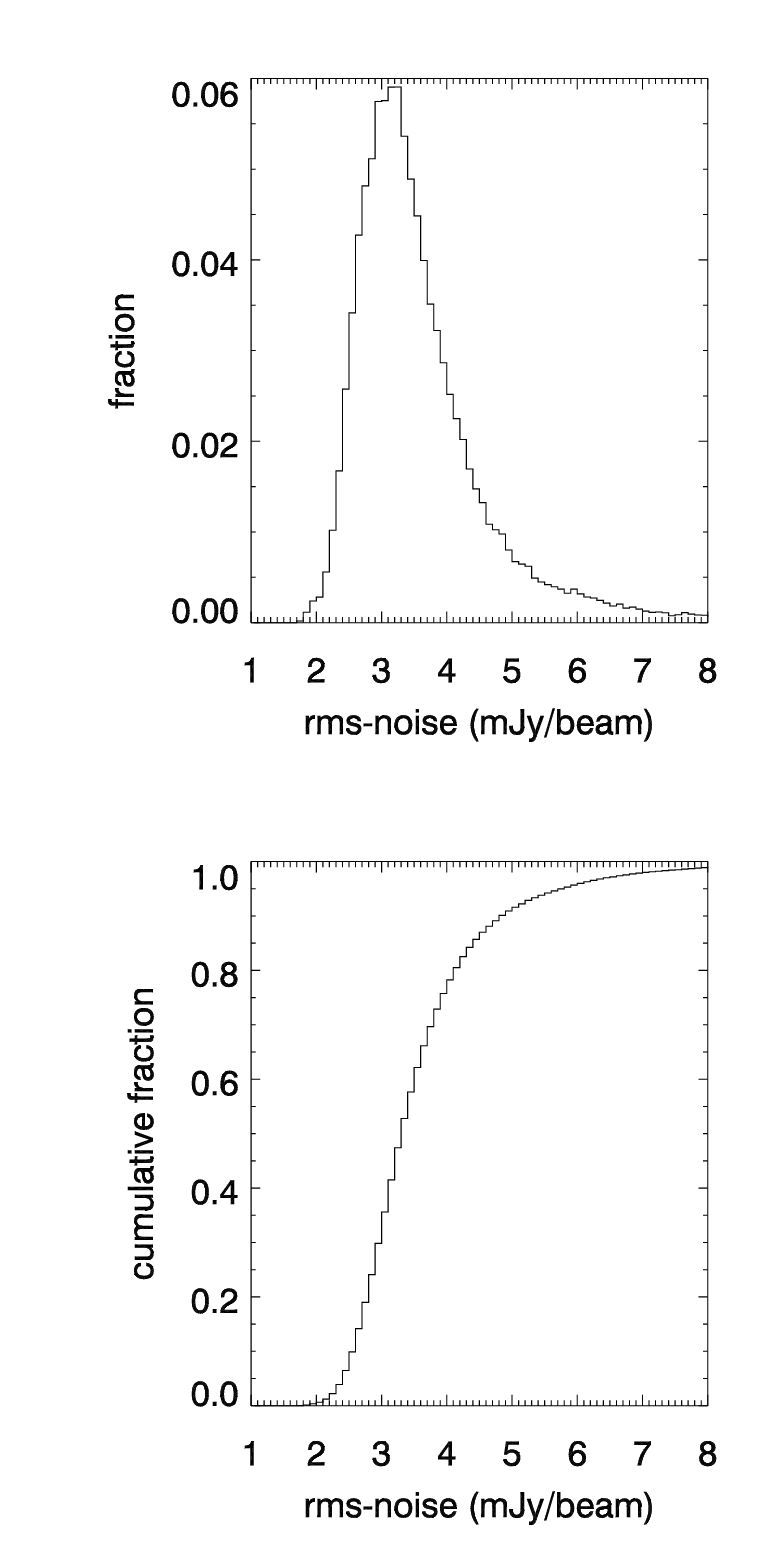}
\caption{The differential (top) and cumulative distribution (bottom) of rms-noise in the WISH.}
\label{noise}
\end{figure}

Figure \ref{noise} shows the distribution of the local noise level in WISH.
The use of the broadband back-end for the WISH leads to slightly lower noise level compared with the main WENSS. However, we do not achieve the noise levels obtained in the WENSS polar cap region, which was observed with the same broadband system. This is probably due to the poorer $u-v$ coverage, resulting in higher sidelobe noise (see also \S 2.4).

\subsubsection{Positional accuracy}
The check the accuracy of the positions in WISH, we want to correlate the WISH with other catalogues with similar or better positional accuracy. Ideally, we would like to use accurate optical catalogues such as the USNO-A2.0, but due to the large WISH beam, there are on average more than one optical counterpart within one wish beam. We therefore used other radio catalogues overlapping with the WISH area.

The only radio catalogues in the WISH area that have position accuracies $<$1\arcsec\ are the TEXAS \citep{dou96} and NVSS \citep{con98}. Using the TEXAS, we find a mean position difference of $-$0\farcs54 (median $-$0\farcs55) in Right Ascension (RA) and $-$0\farcs27 (median $-$0\farcs39) in Declination (DEC).
The NVSS has better position accuracy than the TEXAS, and we have therefore used NVSS sources as position calibrators for WISH. A comparison of the WISH and NVSS positions is therefore not independent, and should be treated with caution. We still perform such a comparison, as our sample of USS sources (\S 3) is based on this correlation.

\begin{figure*}
\centering
\includegraphics[width=16cm]{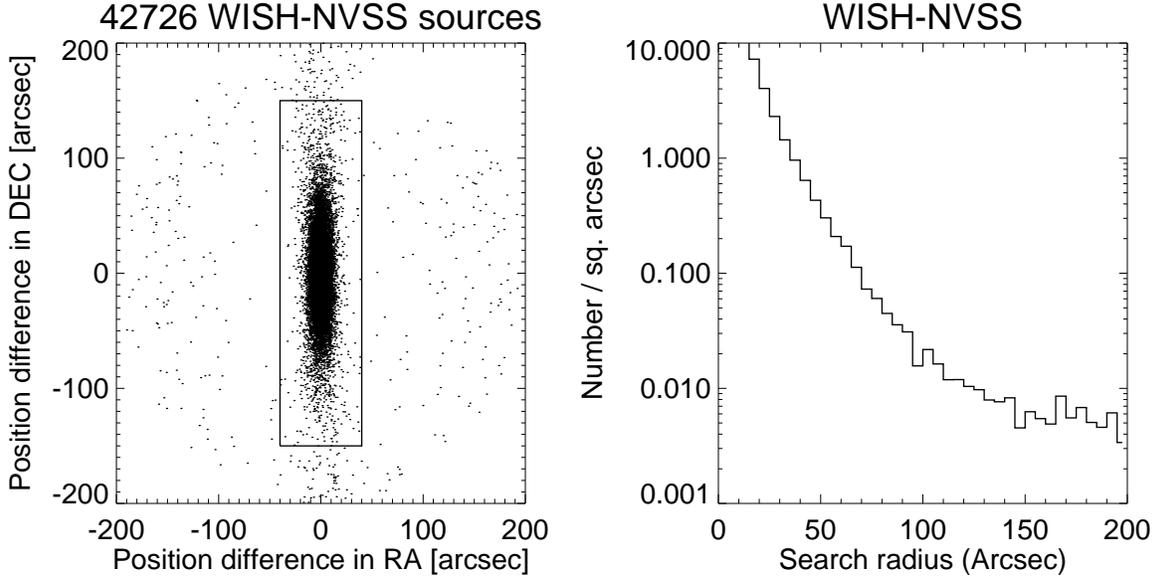}
\caption{{\it Left:} Plot showing the relative position differences between WISH and NVSS. Only single component WISH sources (\S 2.3.3) with a single NVSS counterpart within 200\arcsec\ are considered. Note the strong asymmetry of this distribution due to the very elongated WISH synthesized beam. The box indicates the region where we accept only 1 NVSS source as being associated with a single WISH source (\S 4). {\it Right:} The density of NVSS sources around a WISH source. Note that the distribution flattens from 150\arcsec\ onwards.}
\label{wishnvsspos}
\end{figure*}

To avoid problems due to the different resolution of WISH and NVSS, we only retained WISH sources fitted with a single Gaussian component (see \S 2.3.3), and excluded all objects with $\ge 2$ NVSS sources within 200\arcsec. Figure~\ref{wishnvsspos} shows the relative position differences. The most obvious feature is the strong asymmetry in the position differences due to the very elongated synthesized beam of the WISH. This effect is not seen in the comparison with the TEXAS survey, as this survey groups sources $<$2\arcmin\ as a single entry in the catalogue.

The mean offset between WISH and NVSS is $<\Delta(RA)>=-0\farcs59$ (median $-$0\farcs51) and $<\Delta(DEC)>=-0\farcs32$ (median $-$0\farcs37). These values are consistent with those found from the correlation with the TEXAS. The offset can also be compared with the systematic offset between the WENSS and NVSS positions: $<\Delta(RA)>=-0.12$ and $<\Delta(DEC)>=-0.09$.
However, for consistency with the WENSS, and in order to retain independent positions for comparison with future surveys, we did not apply this correction to the WISH catalogue.

We find a significantly larger offset in RA, which is due to the lack of correction for polar motion and/or a timing problem \citep[see also][]{dev02}.
Despite this large scatter due to the elongated beam, the offset in DEC is only half that in RA, indicating that the large Gaussian fitting errors do not introduce a systematic bias. 

\subsubsection{Morphology}
As can be seen from Fig.~\ref{sampleimage}, WISH has poor resolution in declination. The mean ratio between the fitted major and minor axis $<b_{maj}/b_{min}>=5.00$, while for WENSS, $<b_{maj}/b_{min}>=1.89$. This illustrates the strong difference in ellipticity of the synthesised beam. Because we used the same source finding algorithm as WENSS, this leads to a large number of mis-classifications of WISH sources as resolved North-South sources.
Because the NVSS already covers the same area at 45\arcsec\ resolution, the NVSS morphological parameters should be used in any selection based on such parameters. We include the WISH source dimensions only for consistency with the 2 WENSS catalogues.

\subsection{Flux density}
\subsubsection{Comparison with other radio surveys}
To check the flux calibration of the WISH, we correlated WISH with several other radio surveys. Because no previous 352~MHz survey is available covering this part of the sky, we have to use surveys at different frequencies. These are the 365~MHz Texas survey \citep{dou96}, the 408~MHz MRC \cite{lar81}, and the 1.4~GHz NVSS \citep{con98}.
Both the Texas and MRC surveys use the TXS flux scale, which is related to the more commonly used Baars (BA) scale by the relation $S(TXS)=0.9607 \times S(BA)$ \citep{dou96}. To compare the flux densities, we have put the Texas and MRC flux densities on the Baars scale, as used for WISH.

For the Texas survey, the difference in survey frequency should cause the WISH flux densities to be only $\sim$3\% brighter than the Texas flux densities (assuming $\alpha_{352}^{365}=-0.8$). The observed flux ratio is clearly higher: $<S_{WISH}/S_{Texas}=1.071>$ (median 1.059). A similar result is seen for the MRC: $<S_{WISH}/S_{MRC}=1.242>$ (median 1.262) with an expected value of 1.125. For the NVSS, the flux differences are larger, but they also point towards a $\sim$16\% over-estimation of the WISH flux densities: $<S_{WISH}/S_{NVSS}=3.500>$ (median 3.551) with an expected value of 3.018.

\subsubsection{Cause of the flux density problem}
We examined the dependence of this flux discrepancy on several source parameters such as flux density, morphology, and position difference with respect to the other surveys. None of these appear to influence this flux discrepancy. 

We do find a strong variation of mean spectral index between individual survey frames (Table~\ref{frames}). There is no dependence on Galactic latitude or longitude, or on Galactic extinction, nor is there a dependence on distance from the field centre. There is a clear discontinuity across the frame boundaries, but no strong inter-frame correlations, suggesting the flux discrepancy depends rather on the epoch of observation than on the position on the sky.

The most obvious explication for these large flux density errors are the low elevations at which the WISH has been observed. This could lead to errors in the system temperature corrections, especially when using flux density calibrators observed at much higher elevations.

\subsubsection{Correction using NVSS}
Because none of the other radio surveys have the same frequency as WISH, we can only apply a statistical correction factor to the individual frames, assuming a constant spectral index between the two survey frequencies. 
Although they are close in survey frequency, we prefer not to use the Texas and MRC surveys, because they are much shallower than the WISH, and have $2-3 \times$ poorer resolution. We shall therefore only use the NVSS in the following.

To determine the expected WISH$-$NVSS spectral indices, we use the spectral index distribution of the 13600 sources in the WENSS$-$NVSS polar cap region. 
Here, we consider only the sources from the WENSS polar cap, because they were observed with the same broadband receiver system as WISH\footnote{We have checked that the WENSS polar cap region is not subject to this flux density problem; this implies that the problem is not due to the broadband receiver system itself.}. As noted by \citet{ren99b}, this system leads to bandwidth smearing, which attenuates faint sources, leading to an underestimate of the flux densities of the order of 10\% for the faintest ($S_{352} \simlt 30$~mJy) sources. To avoid this problem, we therefore consider only sources with $S_{352} > 40$~mJy. As shown by \citet{deb00}, this does not significantly affect the spectral index distribution. 
Because the steep part of the WENSS$-$NVSS spectral index distribution has a nearly Gaussian distribution, we use the fitted peak $GP_{\rm WENSS-NVSS}=-0.786$ of the distribution to compare with the WISH$-$NVSS spectral indices. In each WISH frame, there are on average $\sim 550$ single component sources with $S_{352}>40$~mJy and an NVSS counterpart. 
We also fit a Gaussian to their spectral index distribution to determine $GP_{\rm WISH-NVSS}$ for each frame. Forcing these peaks to coincide with $GP_{\rm WENSS-NVSS}$ yields correction factors $$C=(1400/352)^{GP_{\rm WISH-NVSS}-GP_{\rm WENSS-NVSS}}.$$ 
Table~\ref{frames} lists these correction factors for each frame. They range from 0.691 to 1.199, with a mean of 0.868 (median 0.846), indicating the WISH flux densities are statistically overestimated by an average $\sim 13$\%.

We have also determined the correction factors comparing the mean and median spectral WISH$-$NVSS and WENSS$-$NVSS spectral indices. The resulting correction factors are consistent with the ones determined from the Gaussian peaks, with a mean difference of $\sim 1$\%, and a maximum difference of 7\% and 2\% for the correction factors based on the mean and median, respectively. We therefore estimate these statistical correction factors to be accurate to $<$2\%.

We have applied these correction factors to flux densities listed in the WISH catalogue published on the WENSS homepage, and will consider only the corrected values in the remainder of this paper.

\section{USS sample selection}
We now use the WISH to define a sample of USS sources analogous to the WENSS$-$NVSS sample of \citet{deb00}. We use the same selection criteria, i.e. only sources outside the Galactic Plane $|b|>10$\degr\ with $\alpha_{352}^{1400} < -1.30$, $S_{1400} \ge 10$~mJy, and position differences between WISH and NVSS $<$10\arcsec.
We also need to exclude objects which are listed as a single source in WISH, but resolved in the NVSS. This can occur due to differences in resolution and source finding algorithms, and could easily introduce spurious USS sources. In the WENSS$-$NVSS, \citet{deb00} conservatively excluded all WENSS sources with $>$1 NVSS source within a 72\arcsec\ from the WENSS position. 
Because the WISH synthesized beam is much more asymmetrical than the WENSS beam, we cannot use such a circular beam here. Based on Figure~\ref{wishnvsspos}, we adopt a box of 40\arcsec\ in RA and 150\arcsec\ in DEC to exclude the multiple-component sources.
This results in a sample of 154 USS sources, listed in Table ~\ref{uss}.

\section{Discussion}
Although we used almost the same selection criteria, the source density of the WISH$-$NVSS USS sample (78 sr$^{-1}$) is roughly half that of the WENSS$-$NVSS USS sample. This can be explained by the much larger position uncertainty in declination of the WISH survey (\S 2.3.2), which results in a much larger number of sources that fall outside our adopted circular 10\arcsec\ search radius. Contrary to the asymmetrical exclusion box for multiple NVSS sources (\S 3), we keep a circular correlation search radius because the sources with larger position offsets could also have larger flux density uncertainties, and could also select sources with large angular sizes ($\simgt 1\arcmin$), which are less likely to be at very high redshifts. This implicitly introduces a small bias against sources oriented North-South.

We have searched the literature for previous observations of the sources in this USS sample. There is a significant overlap (16 sources) with the shallower Texas$-$NVSS USS sample of \cite{deb00}. This also includes the highest redshift radio galaxy known to date, TN~J0924$-$2201 \citep[$z=5.19$][]{wvb99}. Six sources are also detected in the second data release of the 2MASS survey \citet{jar00}, suggesting they are identified with low redshift galaxy clusters. This is consistent with the samples of \citet{deb00}, who found that $>$3\% of their USS sources are associated with nearby galaxy clusters. 

\section{Conclusions}
The WISH survey is the deepest low-frequency survey covering roughly a quarter of the area between the WENSS and SUMSS surveys ($-30\degr<DEC<+28\degr$). Because it has been observed at very low elevations, it has relatively poor declination resolution compared with the NVSS. It is sufficiently deep to provide spectral index information for $\sim$42000 sources in common with the NVSS. This provides a unique sample of faint USS sources which can be observed with southern hemisphere telescopes. We have obtained VLA and/or ATCA snapshot observations of a first batch of 69 sources from this sample to provide more accurate positions and morphological information needed for the identification of the host galaxies. A campaign of $K-$band identifications with CTIO, and optical spectroscopy with the VLT has already identified several new $z>3$ radio galaxies from this sample (de Vries \etal, in preparation).

\begin{acknowledgements}
We thank Wim de Vries of useful discussions.
The Westerbork Synthesis Radio Telescope (WSRT) is operated by the Netherlands Foundation for Research in Astronomy (NFRA) with financial support of the Netherlands Organization for Scientific Research (NWO).
This work was supported by a Marie Curie Fellowship of the European Community programme 'Improving Human Research Potential and the Socio-Economic Knowledge Base' under contract number HPMF-CT-2000-00721.
The work by W.v.B. at IGPP/LLNL was performed under the auspices of the U.S. Department of Energy, National Nuclear Security Administration by the University of California, Lawrence Livermore National Laboratory under contract No. W-7405-Eng-48.
\end{acknowledgements}

\begin{table*}
\caption{The WISH$-$NVSS USS sample}
\label{uss}
\begin{center}
\begin{tiny}
\begin{tabular}{lrrrrrrrl}\hline
Name & $S_{352}$ & $S_{1400}$ & $\alpha_{352}^{1400}$ & $\alpha_{J2000}$ & $\delta_{J2000}$ & $z$ & ID$^a$ & Reference \\
 & mJy & mJy & & $^h\;\; ^m\;\;\;\; ^s\;\;\,$ & \degr$\;\;\;$ \arcmin$\;\;\;$ \arcsec$\;$ & & \\
\hline
WN~J0015$-$1528 &  104$\pm$ 6 &  15.5$\pm$ 0.7 & $-$1.38$\pm$0.06 & 00 15 59.65 & $-$15 28 20.3 & & & \\
WN~J0017$-$2120 &   85$\pm$ 5 &  13.7$\pm$ 0.6 & $-$1.32$\pm$0.05 & 00 17 48.34 & $-$21 20 54.9 & & & \\
WN~J0036$-$1835 &  101$\pm$ 6 &  16.6$\pm$ 1.0 & $-$1.30$\pm$0.06 & 00 36 54.52 & $-$18 35 48.7 & & & \\
WN~J0037$-$1904 &  422$\pm$18 &  64.9$\pm$ 2.4 & $-$1.36$\pm$0.04 & 00 37 23.76 & $-$19 04 32.2 & & & \\
WN~J0038$-$1540 &  406$\pm$17 &  62.7$\pm$ 1.9 & $-$1.35$\pm$0.04 & 00 38 47.15 & $-$15 40 06.1 & & & \citet{deb00} \\
&&&&&&&& \\
WN~J0138$-$1420 &   82$\pm$ 6 &  10.7$\pm$ 0.6 & $-$1.47$\pm$0.07 & 01 38 30.25 & $-$14 20 07.6 & & & \\
WN~J0141$-$1406 &  296$\pm$13 &  29.3$\pm$ 1.3 & $-$1.68$\pm$0.05 & 01 41 38.02 & $-$14 06 08.8 & & & \\
WN~J0151$-$1417 &  180$\pm$ 9 &  18.4$\pm$ 1.0 & $-$1.65$\pm$0.05 & 01 51 38.85 & $-$14 17 44.7 & & & \\
WN~J0203$-$1315 &  215$\pm$10 &  35.4$\pm$ 1.2 & $-$1.31$\pm$0.04 & 02 03 28.09 & $-$13 15 07.1 & & & \\
WN~J0220$-$2311 &  366$\pm$16 &  58.1$\pm$ 1.8 & $-$1.33$\pm$0.04 & 02 20 24.08 & $-$23 11 39.6 & & & \\
&&&&&&&& \\
WN~J0223$-$1539 &  104$\pm$ 6 &  12.1$\pm$ 0.6 & $-$1.56$\pm$0.05 & 02 23 31.53 & $-$15 39 52.5 & & & \\
WN~J0224$-$1701 &   65$\pm$ 6 &  10.0$\pm$ 0.6 & $-$1.36$\pm$0.08 & 02 24 16.23 & $-$17 01 01.3 & & & \\
WN~J0230$-$2001 &  413$\pm$18 &  55.6$\pm$ 2.1 & $-$1.45$\pm$0.04 & 02 30 45.73 & $-$20 01 18.9 & & & \\
WN~J0230$-$1706 &   64$\pm$ 6 &  10.5$\pm$ 0.6 & $-$1.31$\pm$0.08 & 02 30 52.16 & $-$17 06 26.8 & & & \\
WN~J0246$-$1649 &  368$\pm$15 &  54.9$\pm$ 1.7 & $-$1.38$\pm$0.04 & 02 46 52.89 & $-$16 49 28.1 & & & \\
&&&&&&&& \\
WN~J0306$-$1736 &  125$\pm$ 6 &  19.8$\pm$ 0.8 & $-$1.34$\pm$0.05 & 03 06 29.62 & $-$17 36 38.6 & & & \\
WN~J0314$-$1849 &  143$\pm$ 8 &  21.2$\pm$ 0.8 & $-$1.38$\pm$0.05 & 03 14 13.66 & $-$18 49 50.7 & & & \\
WN~J0316$-$2137 &  100$\pm$ 6 &  15.6$\pm$ 0.7 & $-$1.34$\pm$0.06 & 03 16 21.01 & $-$21 37 41.3 & & & \\
WN~J0330$-$1810 &   96$\pm$ 6 &  14.9$\pm$ 0.6 & $-$1.35$\pm$0.05 & 03 30 50.47 & $-$18 10 59.6 & & & \\
WN~J0335$-$2041 &  364$\pm$16 &  53.5$\pm$ 2.0 & $-$1.39$\pm$0.04 & 03 35 40.08 & $-$20 41 10.4 & & & \\
&&&&&&&& \\
WN~J0340$-$2159 &  313$\pm$14 &  51.6$\pm$ 2.0 & $-$1.31$\pm$0.04 & 03 40 14.93 & $-$21 59 54.4 & & & \\
WN~J0341$-$1719 &  191$\pm$ 9 &  27.6$\pm$ 1.3 & $-$1.40$\pm$0.05 & 03 41 55.52 & $-$17 19 27.0 & & & \\
WN~J0349$-$1801 &   94$\pm$ 6 &  15.6$\pm$ 0.7 & $-$1.30$\pm$0.06 & 03 49 25.11 & $-$18 01 31.7 & & & \\
WN~J0414$-$2114 &  307$\pm$13 &  47.0$\pm$ 1.5 & $-$1.36$\pm$0.04 & 04 14 01.22 & $-$21 14 53.9 & & & \\
WN~J0423$-$1537 &  217$\pm$ 9 &  31.8$\pm$ 1.7 & $-$1.39$\pm$0.05 & 04 23 32.85 & $-$15 37 43.1 & & & \\
&&&&&&&& \\
WN~J0456$-$2202 &   66$\pm$ 5 &  10.9$\pm$ 0.6 & $-$1.31$\pm$0.07 & 04 56 49.41 & $-$22 02 55.5 & & & \\
WN~J0510$-$1838 & 6217$\pm$254& 634.0$\pm$20.7 & $-$1.65$\pm$0.04 & 05 10 32.43 & $-$18 38 42.5 & & 2MASS & \citet{jar00} \\
WN~J0526$-$1830 &  134$\pm$ 6 &  19.6$\pm$ 1.3 & $-$1.39$\pm$0.06 & 05 26 24.60 & $-$18 30 40.1 & & & \\
WN~J0526$-$2145 &  112$\pm$ 6 &  17.6$\pm$ 0.7 & $-$1.34$\pm$0.05 & 05 26 48.80 & $-$21 45 19.8 & & & \\
WN~J0528$-$1710 &  131$\pm$ 7 &  21.8$\pm$ 0.8 & $-$1.30$\pm$0.05 & 05 28 04.93 & $-$17 10 04.1 & & & \\
&&&&&&&& \\
WN~J0536$-$1357 &  279$\pm$14 &  40.1$\pm$ 1.6 & $-$1.41$\pm$0.05 & 05 36 51.96 & $-$13 57 10.3 & & 2MASS & \citet{jar00} \\
WN~J0557$-$1124 &  351$\pm$15 &  40.1$\pm$ 1.9 & $-$1.57$\pm$0.05 & 05 57 00.72 & $-$11 24 16.6 & & & \\
WN~J0602$-$2036 &  235$\pm$11 &  37.7$\pm$ 1.2 & $-$1.33$\pm$0.04 & 06 02 29.41 & $-$20 36 46.3 & & & \\
WN~J0604$-$2015 &  222$\pm$10 &  27.6$\pm$ 1.0 & $-$1.51$\pm$0.04 & 06 04 57.58 & $-$20 15 56.9 & & & \\
WN~J0621$-$1902 &   64$\pm$ 5 &  10.3$\pm$ 1.0 & $-$1.32$\pm$0.09 & 06 21 05.78 & $-$19 02 03.6 & & & \\
&&&&&&&& \\
WN~J0851$-$1728 &  101$\pm$ 6 &  15.1$\pm$ 0.7 & $-$1.38$\pm$0.05 & 08 51 25.09 & $-$17 28 43.8 & & & \\
WN~J0903$-$1759 &  243$\pm$11 &  38.2$\pm$ 1.2 & $-$1.34$\pm$0.04 & 09 03 44.11 & $-$17 59 52.5 & & & \\
WN~J0910$-$2228 &  469$\pm$20 &  53.5$\pm$ 1.7 & $-$1.57$\pm$0.04 & 09 10 34.15 & $-$22 28 43.3 & & $R$ & \citet{deb00} \\
WN~J0912$-$1655 &  104$\pm$ 7 &  11.8$\pm$ 0.6 & $-$1.58$\pm$0.06 & 09 12 57.24 & $-$16 55 54.8 & & & \\
WN~J0924$-$2201 &  454$\pm$19 &  71.1$\pm$ 2.2 & $-$1.34$\pm$0.04 & 09 24 19.94 & $-$22 01 42.2 & 5.19 & $R,K$ & \citet{wvb99} \\
&&&&&&&& \\
WN~J0938$-$1831 &  116$\pm$ 7 &  18.4$\pm$ 1.2 & $-$1.33$\pm$0.06 & 09 38 44.24 & $-$18 31 36.2 & & & \\
WN~J0945$-$2118 &  160$\pm$ 8 &  18.3$\pm$ 0.7 & $-$1.57$\pm$0.04 & 09 45 29.43 & $-$21 18 48.8 & & DSS & \\
WN~J0949$-$2520 &  103$\pm$ 8 &  15.9$\pm$ 0.7 & $-$1.35$\pm$0.07 & 09 49 28.04 & $-$25 20 13.4 & & & \\
WN~J0956$-$1500 &  440$\pm$19 &  67.9$\pm$ 2.1 & $-$1.35$\pm$0.04 & 09 56 01.46 & $-$15 00 21.5 & & & \citet{deb00} \\
WN~J1001$-$1723 &   70$\pm$ 5 &  10.8$\pm$ 0.6 & $-$1.36$\pm$0.07 & 10 01 15.75 & $-$17 23 27.2 & & & \\
&&&&&&&& \\
WN~J1014$-$1830 &   86$\pm$ 6 &  11.6$\pm$ 0.6 & $-$1.45$\pm$0.06 & 10 14 10.48 & $-$18 30 00.9 & & DSS & \\
WN~J1026$-$2116 &  444$\pm$19 &  60.2$\pm$ 1.9 & $-$1.45$\pm$0.04 & 10 26 22.35 & $-$21 16 07.8 & & $R,K$ & \citet{deb00} \\
WN~J1036$-$1837 &  446$\pm$19 &  70.8$\pm$ 2.2 & $-$1.33$\pm$0.04 & 10 36 16.19 & $-$18 37 28.0 & & & \\
WN~J1047$-$1836 &  153$\pm$ 8 &  20.4$\pm$ 0.8 & $-$1.46$\pm$0.05 & 10 47 15.51 & $-$18 36 31.1 & & & \\
WN~J1051$-$1517 &  103$\pm$ 8 &  16.6$\pm$ 0.7 & $-$1.32$\pm$0.06 & 10 51 45.62 & $-$15 17 06.3 & & & \\
&&&&&&&& \\
WN~J1052$-$1812 &  118$\pm$ 7 &  14.4$\pm$ 0.6 & $-$1.52$\pm$0.05 & 10 52 00.82 & $-$18 12 32.3 & & & \\
WN~J1053$-$1656 &  226$\pm$10 &  35.8$\pm$ 1.5 & $-$1.34$\pm$0.05 & 10 53 19.77 & $-$16 56 40.2 & & & \\
WN~J1103$-$2004 &  189$\pm$ 9 &  30.7$\pm$ 1.0 & $-$1.32$\pm$0.04 & 11 03 17.51 & $-$20 04 58.9 & & & \\
WN~J1104$-$1913 &  203$\pm$10 &  29.3$\pm$ 1.4 & $-$1.40$\pm$0.05 & 11 04 07.25 & $-$19 13 47.6 & & DSS & \\
WN~J1109$-$1917 & 1324$\pm$54 & 197.4$\pm$ 5.9 & $-$1.38$\pm$0.04 & 11 09 49.93 & $-$19 17 53.7 & & & \\
&&&&&&&& \\
WN~J1123$-$2154 &  299$\pm$13 &  49.3$\pm$ 1.6 & $-$1.30$\pm$0.04 & 11 23 10.11 & $-$21 54 05.6 & & & \\
WN~J1127$-$2126 &   81$\pm$ 6 &  11.8$\pm$ 0.6 & $-$1.40$\pm$0.06 & 11 27 54.80 & $-$21 26 21.8 & & 2MASS & \citet{jar00} \\
WN~J1132$-$2102 &  219$\pm$10 &  30.8$\pm$ 1.0 & $-$1.42$\pm$0.04 & 11 32 52.66 & $-$21 02 45.0 & & & \\
WN~J1138$-$1324 &   83$\pm$ 8 &  10.0$\pm$ 0.6 & $-$1.53$\pm$0.08 & 11 38 05.42 & $-$13 24 23.5 & & & \\
WN~J1143$-$2143 &  423$\pm$18 &  67.6$\pm$ 2.5 & $-$1.33$\pm$0.04 & 11 43 17.43 & $-$21 43 31.2 & & & \\
&&&&&&&& \\
WN~J1148$-$2114 &  101$\pm$ 6 &  15.5$\pm$ 0.7 & $-$1.36$\pm$0.05 & 11 48 13.50 & $-$21 14 03.7 & & & \\
WN~J1150$-$1317 &  205$\pm$ 9 &  31.1$\pm$ 1.0 & $-$1.37$\pm$0.04 & 11 50 09.59 & $-$13 17 53.9 & & & \\
WN~J1151$-$2547 &   86$\pm$ 9 &  11.8$\pm$ 1.0 & $-$1.44$\pm$0.10 & 11 51 46.42 & $-$25 47 52.2 & & DSS & \\
WN~J1152$-$1558 &  116$\pm$ 6 &  14.1$\pm$ 1.5 & $-$1.52$\pm$0.09 & 11 52 50.03 & $-$15 58 00.9 & & & \\
WN~J1200$-$1125 &   74$\pm$ 5 &  10.2$\pm$ 0.6 & $-$1.44$\pm$0.06 & 12 00 54.27 & $-$11 25 48.6 & & & \\
&&&&&&&& \\
WN~J1211$-$1126 &   91$\pm$ 5 &  14.0$\pm$ 1.3 & $-$1.36$\pm$0.08 & 12 11 32.05 & $-$11 26 01.5 & & & \\
WN~J1222$-$2129 &  101$\pm$ 7 &  14.3$\pm$ 0.6 & $-$1.42$\pm$0.06 & 12 22 48.22 & $-$21 29 10.0 & & & \\
WN~J1222$-$2531 &   99$\pm$13 &  11.6$\pm$ 0.6 & $-$1.55$\pm$0.10 & 12 22 50.24 & $-$25 31 18.0 & & 2MASS & \citet{abe89}\\
WN~J1224$-$2101 &  103$\pm$ 6 &  16.9$\pm$ 0.7 & $-$1.31$\pm$0.05 & 12 24 30.26 & $-$21 01 41.9 & & & \\
WN~J1225$-$1429 &  245$\pm$11 &  31.2$\pm$ 1.0 & $-$1.49$\pm$0.04 & 12 25 31.71 & $-$14 29 15.4 & & & \\
\end{tabular}
\end{tiny}
\end{center}
\end{table*}

\begin{table*}
\begin{center}
\begin{tiny}
\begin{tabular}{lrrrrrrrl}\hline
Name & $S_{352}$ & $S_{1400}$ & $\alpha_{352}^{1400}$ & $\alpha_{J2000}$ & $\delta_{J2000}$ & $z$ & ID & Reference \\
 & mJy & mJy & & $^h\;\; ^m\;\;\;\; ^s\;\;\,$ & \degr$\;\;\;$ \arcmin$\;\;\;$ \arcsec$\;$ & & \\
\hline
WN~J1230$-$1038 &   65$\pm$ 6 &  10.3$\pm$ 0.6 & $-$1.33$\pm$0.08 & 12 30 00.61 & $-$10 38 58.5 & & & \\
WN~J1242$-$1616 &  196$\pm$ 9 &  30.9$\pm$ 1.0 & $-$1.34$\pm$0.04 & 12 42 13.34 & $-$16 16 57.4 & & & \\
WN~J1242$-$1646 &  119$\pm$ 7 &  19.7$\pm$ 0.8 & $-$1.30$\pm$0.05 & 12 42 27.56 & $-$16 46 36.4 & & & \\
WN~J1246$-$1245 &  166$\pm$ 8 &  27.4$\pm$ 1.0 & $-$1.30$\pm$0.04 & 12 46 17.32 & $-$12 45 24.6 & & & \\
WN~J1246$-$2426 &  112$\pm$ 7 &  13.8$\pm$ 0.6 & $-$1.52$\pm$0.05 & 12 46 44.15 & $-$24 26 17.8 & & & \\
&&&&&&&& \\
WN~J1252$-$2057 &   70$\pm$ 6 &  11.3$\pm$ 0.6 & $-$1.32$\pm$0.07 & 12 52 02.13 & $-$20 57 12.0 & & & \\
WN~J1252$-$2015 &   99$\pm$ 6 &  15.9$\pm$ 1.0 & $-$1.33$\pm$0.06 & 12 52 50.30 & $-$20 15 02.8 & & & \\
WN~J1254$-$1604 &   84$\pm$ 6 &  12.8$\pm$ 0.6 & $-$1.36$\pm$0.06 & 12 54 11.90 & $-$16 04 44.2 & & & \\
WN~J1255$-$1256 &  184$\pm$ 9 &  25.1$\pm$ 0.9 & $-$1.44$\pm$0.04 & 12 55 33.48 & $-$12 56 37.0 & & & \\
WN~J1255$-$1913 &  108$\pm$ 6 &  10.8$\pm$ 0.6 & $-$1.67$\pm$0.06 & 12 55 52.66 & $-$19 13 00.6 & & & \\
&&&&&&&& \\
WN~J1255$-$1222 & 2292$\pm$94 & 321.6$\pm$ 9.7 & $-$1.42$\pm$0.04 & 12 55 55.45 & $-$12 22 54.7 & & & \\
WN~J1300$-$1146 &  317$\pm$15 &  48.7$\pm$ 1.5 & $-$1.36$\pm$0.04 & 13 00 46.89 & $-$11 46 22.8 & & & \\
WN~J1305$-$1441 &  121$\pm$ 7 &  19.8$\pm$ 0.8 & $-$1.31$\pm$0.05 & 13 05 56.83 & $-$14 41 57.9 & & & \\
WN~J1315$-$1511 &   67$\pm$ 7 &  10.2$\pm$ 0.6 & $-$1.36$\pm$0.09 & 13 15 43.95 & $-$15 11 53.9 & & & \\
WN~J1326$-$2330 &  523$\pm$22 &  80.2$\pm$ 2.5 & $-$1.36$\pm$0.04 & 13 26 25.24 & $-$23 30 23.4 & & DSS & \citet{deb00} \\
&&&&&&&& \\
WN~J1331$-$1947 &  488$\pm$20 &  70.8$\pm$ 2.2 & $-$1.40$\pm$0.04 & 13 31 47.18 & $-$19 47 26.5 & & & \\
WN~J1345$-$2423 &  130$\pm$ 9 &  19.1$\pm$ 0.8 & $-$1.39$\pm$0.06 & 13 45 47.68 & $-$24 23 45.6 & & & \\
WN~J1349$-$2049 &  203$\pm$10 &  25.4$\pm$ 0.9 & $-$1.51$\pm$0.04 & 13 49 02.91 & $-$20 49 11.1 & & DSS & \\
WN~J1351$-$1205 &  183$\pm$ 9 &  19.6$\pm$ 1.2 & $-$1.62$\pm$0.06 & 13 51 14.67 & $-$12 05 53.6 & & & \\
WN~J1353$-$2152 &  347$\pm$15 &  46.0$\pm$ 1.8 & $-$1.46$\pm$0.04 & 13 53 55.92 & $-$21 52 12.6 & & & \\
&&&&&&&& \\
WN~J1416$-$1130 &  912$\pm$37 & 142.0$\pm$ 5.5 & $-$1.35$\pm$0.04 & 14 16 04.09 & $-$11 30 30.2 & & & \\
WN~J1417$-$1522 &   89$\pm$10 &  14.3$\pm$ 0.7 & $-$1.32$\pm$0.09 & 14 17 29.38 & $-$15 22 30.0 & & & \\
WN~J1429$-$1909 &   99$\pm$ 7 &  14.8$\pm$ 0.6 & $-$1.37$\pm$0.06 & 14 29 07.73 & $-$19 09 50.5 & & & \\
WN~J1437$-$2040 &  118$\pm$ 7 &  12.5$\pm$ 1.0 & $-$1.63$\pm$0.07 & 14 37 06.06 & $-$20 40 21.0 & & & \\
WN~J1446$-$2003 & 1164$\pm$48 & 182.1$\pm$ 5.5 & $-$1.34$\pm$0.04 & 14 46 48.51 & $-$20 03 37.7 & 0.753 & $R$ & \citet{gir96} \\
&&&&&&&& \\
WN~J1451$-$2301 &  128$\pm$ 7 &  14.6$\pm$ 1.2 & $-$1.57$\pm$0.07 & 14 51 15.54 & $-$23 01 46.5 & & & \\
WN~J1454$-$1130 &   76$\pm$ 6 &  10.2$\pm$ 1.1 & $-$1.46$\pm$0.09 & 14 54 49.59 & $-$11 30 03.0 & & & \\
WN~J1457$-$0954 &  120$\pm$ 9 &  15.5$\pm$ 1.4 & $-$1.48$\pm$0.09 & 14 57 53.71 & $-$09 54 22.9 & & & \\
WN~J1513$-$1801 &  544$\pm$23 &  83.0$\pm$ 2.5 & $-$1.36$\pm$0.04 & 15 13 55.31 & $-$18 01 07.9 & & & \citet{deb00} \\
WN~J1516$-$2110 &  230$\pm$10 &  34.1$\pm$ 1.1 & $-$1.38$\pm$0.04 & 15 16 42.32 & $-$21 10 27.4 & & & \\
&&&&&&&& \\
WN~J1518$-$1225 &  105$\pm$ 6 &  10.5$\pm$ 0.6 & $-$1.67$\pm$0.06 & 15 18 43.43 & $-$12 25 35.6 & & & \\
WN~J1544$-$2104 &  243$\pm$11 &  24.3$\pm$ 1.2 & $-$1.67$\pm$0.05 & 15 44 04.77 & $-$21 04 11.2 & & & \\
WN~J1551$-$1422 &   83$\pm$ 5 &  10.7$\pm$ 0.6 & $-$1.48$\pm$0.06 & 15 51 58.78 & $-$14 22 48.3 & & & \\
WN~J1557$-$1349 &   90$\pm$ 6 &  13.3$\pm$ 0.6 & $-$1.39$\pm$0.06 & 15 57 41.72 & $-$13 49 54.8 & & & \\
WN~J1558$-$1142 & 1316$\pm$54 & 201.8$\pm$ 6.1 & $-$1.36$\pm$0.04 & 15 58 00.63 & $-$11 42 24.9 & & & \\
&&&&&&&& \\
WN~J1603$-$1500 &  127$\pm$ 7 &  17.4$\pm$ 0.7 & $-$1.44$\pm$0.05 & 16 03 04.78 & $-$15 00 53.8 & & & \\
WN~J1605$-$1952 &  241$\pm$11 &  39.2$\pm$ 1.3 & $-$1.32$\pm$0.04 & 16 05 53.98 & $-$19 52 00.6 & & & \\
WN~J1631$-$1332 &   81$\pm$ 5 &  12.6$\pm$ 0.6 & $-$1.35$\pm$0.06 & 16 31 49.83 & $-$13 32 28.7 & & & \\
WN~J1633$-$1517 &   93$\pm$ 7 &  12.0$\pm$ 0.6 & $-$1.48$\pm$0.06 & 16 33 15.09 & $-$15 17 25.9 & & & \\
WN~J1634$-$1107 &  172$\pm$ 9 &  17.4$\pm$ 0.7 & $-$1.66$\pm$0.05 & 16 34 41.37 & $-$11 07 12.9 & & & \\
&&&&&&&& \\
WN~J1637$-$1931 &  327$\pm$14 &  35.8$\pm$ 1.2 & $-$1.60$\pm$0.04 & 16 37 44.85 & $-$19 31 22.9 & & & \citet{deb00} \\
WN~J1653$-$1756 &   73$\pm$ 5 &  10.1$\pm$ 1.2 & $-$1.44$\pm$0.10 & 16 53 50.11 & $-$17 56 04.6 & & & \\
WN~J1653$-$1155 & 1179$\pm$48 & 191.5$\pm$ 5.8 & $-$1.32$\pm$0.04 & 16 53 52.81 & $-$11 55 59.1 & & & \citet{deb00} \\
WN~J1702$-$1414 &  156$\pm$11 &  25.0$\pm$ 0.9 & $-$1.33$\pm$0.06 & 17 02 51.45 & $-$14 14 55.8 & & & \\
WN~J1932$-$1931 & 4841$\pm$198& 789.4$\pm$23.7 & $-$1.31$\pm$0.04 & 19 32 07.22 & $-$19 31 49.6 & & & \citet{deb00} \\
&&&&&&&& \\
WN~J1939$-$1457 &  193$\pm$11 &  28.8$\pm$ 1.0 & $-$1.38$\pm$0.05 & 19 39 13.18 & $-$14 57 27.3 & & & \\
WN~J1942$-$1514 &  385$\pm$22 &  33.5$\pm$ 1.1 & $-$1.77$\pm$0.05 & 19 42 07.50 & $-$15 14 37.9 & & & \\
WN~J1954$-$1207 &  636$\pm$27 &  98.9$\pm$ 3.0 & $-$1.35$\pm$0.04 & 19 54 24.24 & $-$12 07 49.3 & & $K$ & \citet{deb00} \\
WN~J1956$-$1308 &   96$\pm$ 6 &  15.9$\pm$ 0.7 & $-$1.31$\pm$0.06 & 19 56 06.62 & $-$13 08 08.4 & & & \\
WN~J2002$-$1842 &   81$\pm$ 5 &  11.4$\pm$ 0.6 & $-$1.42$\pm$0.06 & 20 02 56.00 & $-$18 42 47.8 & & & \\
&&&&&&&& \\
WN~J2007$-$1840 &  230$\pm$10 &  34.6$\pm$ 1.1 & $-$1.37$\pm$0.04 & 20 07 00.16 & $-$18 40 57.3 & & & \\
WN~J2007$-$1843 &  102$\pm$ 5 &  16.7$\pm$ 1.0 & $-$1.31$\pm$0.06 & 20 07 31.49 & $-$18 43 43.9 & & & \\
WN~J2007$-$1316 &  922$\pm$38 & 112.9$\pm$ 3.4 & $-$1.52$\pm$0.04 & 20 07 53.23 & $-$13 16 45.0 & & $K$ & \citet{deb00} \\
WN~J2014$-$2115 &  311$\pm$14 &  47.3$\pm$ 1.5 & $-$1.36$\pm$0.04 & 20 14 31.69 & $-$21 15 02.6 & & & \citet{deb00} \\
WN~J2027$-$1909 &   86$\pm$ 5 &  11.7$\pm$ 0.6 & $-$1.44$\pm$0.06 & 20 27 01.50 & $-$19 09 53.8 & & & \\
&&&&&&&& \\
WN~J2045$-$1948 &  103$\pm$ 6 &  15.5$\pm$ 0.7 & $-$1.37$\pm$0.05 & 20 45 56.24 & $-$19 48 19.6 & & & \\
WN~J2052$-$2306 &  356$\pm$15 &  57.0$\pm$ 2.1 & $-$1.33$\pm$0.04 & 20 52 50.08 & $-$23 06 22.8 & & & \\
WN~J2054$-$2006 &   71$\pm$ 6 &  10.4$\pm$ 0.6 & $-$1.39$\pm$0.08 & 20 54 35.75 & $-$20 06 09.4 & & & \\
WN~J2054$-$1939 &   74$\pm$ 6 &  12.2$\pm$ 0.6 & $-$1.31$\pm$0.07 & 20 54 52.30 & $-$19 39 52.6 & & & \\
WN~J2103$-$1917 & 1358$\pm$56 & 223.8$\pm$ 6.7 & $-$1.31$\pm$0.04 & 21 03 42.91 & $-$19 17 47.5 & & & \citet{deb00} \\
&&&&&&&& \\
WN~J2104$-$2037 &  183$\pm$ 8 &  27.6$\pm$ 0.9 & $-$1.37$\pm$0.04 & 21 04 13.33 & $-$20 37 27.7 & & & \\
WN~J2105$-$1057 &  136$\pm$ 8 &  16.9$\pm$ 0.7 & $-$1.51$\pm$0.05 & 21 05 45.16 & $-$10 57 33.5 & & & \\
WN~J2106$-$1040 &  322$\pm$14 &  50.0$\pm$ 1.9 & $-$1.35$\pm$0.04 & 21 06 39.96 & $-$10 40 43.5 & & DSS & \\
WN~J2114$-$2127 &  203$\pm$ 9 &  33.1$\pm$ 1.1 & $-$1.31$\pm$0.04 & 21 14 26.43 & $-$21 27 50.9 & & & \\
WN~J2116$-$1519 &   73$\pm$11 &  11.0$\pm$ 0.9 & $-$1.37$\pm$0.12 & 21 16 13.07 & $-$15 19 38.6 & & & \\
&&&&&&&& \\
WN~J2133$-$1656 &  234$\pm$10 &  28.3$\pm$ 1.3 & $-$1.53$\pm$0.05 & 21 33 24.84 & $-$16 56 22.2 & & & \\
WN~J2137$-$1246 &   71$\pm$ 6 &  11.2$\pm$ 0.6 & $-$1.34$\pm$0.07 & 21 37 30.62 & $-$12 46 43.7 & & & \\
WN~J2137$-$1350 &  258$\pm$12 &  35.6$\pm$ 1.2 & $-$1.43$\pm$0.04 & 21 37 47.42 & $-$13 50 28.7 & & & \\
WN~J2139$-$1205 &  500$\pm$21 &  82.0$\pm$ 2.5 & $-$1.31$\pm$0.04 & 21 39 58.18 & $-$12 05 28.6 & & & \\
WN~J2141$-$1425 &  114$\pm$ 9 &  18.9$\pm$ 0.7 & $-$1.30$\pm$0.06 & 21 41 18.88 & $-$14 25 48.1 & & & \\
\end{tabular}
\end{tiny}
\end{center}
\end{table*}

\begin{table*}
\begin{center}
\begin{tiny}
\begin{tabular}{lrrrrrrrl}\hline
Name & $S_{352}$ & $S_{1400}$ & $\alpha_{352}^{1400}$ & $\alpha_{J2000}$ & $\delta_{J2000}$ & $z$ & ID & Reference \\
 & mJy & mJy & & $^h\;\; ^m\;\;\;\; ^s\;\;\,$ & \degr$\;\;\;$ \arcmin$\;\;\;$ \arcsec$\;$ & & \\
\hline
WN~J2144$-$1818 &  201$\pm$ 9 &  30.9$\pm$ 1.3 & $-$1.36$\pm$0.05 & 21 44 18.91 & $-$18 18 36.6 & & & \\
WN~J2145$-$2240 &  192$\pm$ 9 &  24.3$\pm$ 1.2 & $-$1.50$\pm$0.05 & 21 45 15.94 & $-$22 40 26.4 & & & \\
WN~J2204$-$1004 &  220$\pm$12 &  28.1$\pm$ 0.9 & $-$1.49$\pm$0.05 & 22 04 00.00 & $-$10 04 21.3 & & & \\
WN~J2214$-$2353 &  188$\pm$ 9 &  29.6$\pm$ 1.3 & $-$1.34$\pm$0.05 & 22 14 14.03 & $-$23 53 25.6 & & & \\
WN~J2216$-$1725 &  458$\pm$20 &  13.4$\pm$ 1.1 & $-$2.56$\pm$0.07 & 22 16 57.57 & $-$17 25 21.4 & 0.1301 & 2MASS & \citet{jar00} \\
&&&&&&&& \\
WN~J2217$-$1913 &  284$\pm$12 &  39.2$\pm$ 1.3 & $-$1.43$\pm$0.04 & 22 17 28.22 & $-$19 13 20.9 & & & \citet{deb00} \\
WN~J2226$-$1151 &   77$\pm$ 5 &  12.5$\pm$ 0.6 & $-$1.31$\pm$0.06 & 22 26 04.17 & $-$11 51 02.3 & & & \\
WN~J2244$-$2520 &   84$\pm$ 8 &  11.5$\pm$ 0.6 & $-$1.44$\pm$0.08 & 22 44 28.75 & $-$25 20 45.3 & & & \\
WN~J2302$-$2229 &  229$\pm$10 &  37.9$\pm$ 1.2 & $-$1.30$\pm$0.04 & 23 02 49.64 & $-$22 29 44.4 & & & \\
WN~J2308$-$1546 &  104$\pm$ 6 &  16.2$\pm$ 0.7 & $-$1.35$\pm$0.05 & 23 08 16.46 & $-$15 46 30.3 & & & \\
&&&&&&&& \\
WN~J2320$-$1436 &   78$\pm$ 6 &  10.1$\pm$ 0.6 & $-$1.48$\pm$0.07 & 23 20 22.54 & $-$14 36 19.6 & & & \\
WN~J2349$-$1542 &  129$\pm$ 7 &  17.2$\pm$ 1.2 & $-$1.46$\pm$0.06 & 23 49 40.09 & $-$15 42 51.3 & & & \\
WN~J2350$-$1321 &  388$\pm$16 &  55.1$\pm$ 1.7 & $-$1.41$\pm$0.04 & 23 50 15.92 & $-$13 21 15.9 & & & \citet{deb00} \\
WN~J2350$-$1238 &  104$\pm$ 6 &  10.5$\pm$ 0.6 & $-$1.66$\pm$0.06 & 23 50 49.70 & $-$12 38 33.1 & & 2MASS & \citet{jar00} \\
\hline
\end{tabular}
\end{tiny}
\end{center}
$^a$ We provide either the name of the large sky survey, or the filter of literature observations.
\end{table*}
\end{document}